\documentclass[a4paper]{jpconf}
\usepackage{graphicx}
\usepackage[utf8]{inputenc}
\usepackage{amsmath}
\usepackage{amssymb}
\usepackage{url}
\usepackage{multicol}
\usepackage{multirow}
\usepackage{array}
\newcolumntype{C}[1]{>{\centering\let\newline\\\arraybackslash\hspace{0pt}}m{#1}}
\newcolumntype{M}[1]{>{\centering\arraybackslash}m{#1}}
\usepackage{citesort}

\begin{document}
\title{Simulation of Nuclear Recoils due to Supernova Neutrino-induced Neutrons in Liquid Xenon Detectors}

\author{Sayan Ghosh$^{1,*}$, Abhijit Bandyopadhyay$^2$, Pijushpani Bhattacharjee$^1$, Sovan Chakraborty$^3$, Kamales Kar$^2$ and Satyajit Saha$^1$}
\address{$^1$Saha Institute of Nuclear Phjysics, HBNI, 1/AF Bidhannagar, Kolkata-700~064, India.}
\address{$^2$Ramakrishna Mission Vivekananda Educational and Research Institute, Belur Math, Howrah 711202, India.}
\address{$^3$Department of Physics, Indian Institute of Technology - Guwahati, Guwahati 781039, India.}
\ead{$^*$sayan.ghosh@saha.ac.in}

\begin{abstract}

    Neutrinos from supernova (SN) bursts can give rise to detectable number of nuclear recoil (NR) events through the coherent elastic neutrino-nucleus scattering (CE$\nu$NS) process in large scale liquid xenon detectors designed for direct dark matter search, depending on the SN progenitor mass and distance. Here we show that in addition to the direct NR events due to CE$\nu$NS process, the SN neutrinos can give rise to additional nuclear recoils due to the elastic scattering of neutrons produced through  inelastic interaction of the neutrinos with the xenon nuclei. We find that the contribution of the supernova neutrino-induced neutrons ($\nu$I$n$) can significantly modify the total xenon NR spectrum at large recoil energies  compared to that expected from the CE$\nu$NS process alone. Moreover, for recoil energies $\gtrsim20$ keV, dominant contribution is obtained from the ($\nu$I$n$) events. We numerically calculate the observable S1 and S2 signals due to both CE$\nu$NS and $\nu$I$n$ processes for a typical liquid xenon based detector, accounting for the multiple scattering effects of the neutrons in the case of $\nu$I$n$, and find that sufficiently large signal  events, those with S1$\gtrsim$50 photo-electrons (PE) and S2$\gtrsim$2300 PE, come mainly from the $\nu$I$n$ scatterings. 
\end{abstract}

\section{Introduction}
\begin{figure}[h]
\centering
\includegraphics[width=18pc]{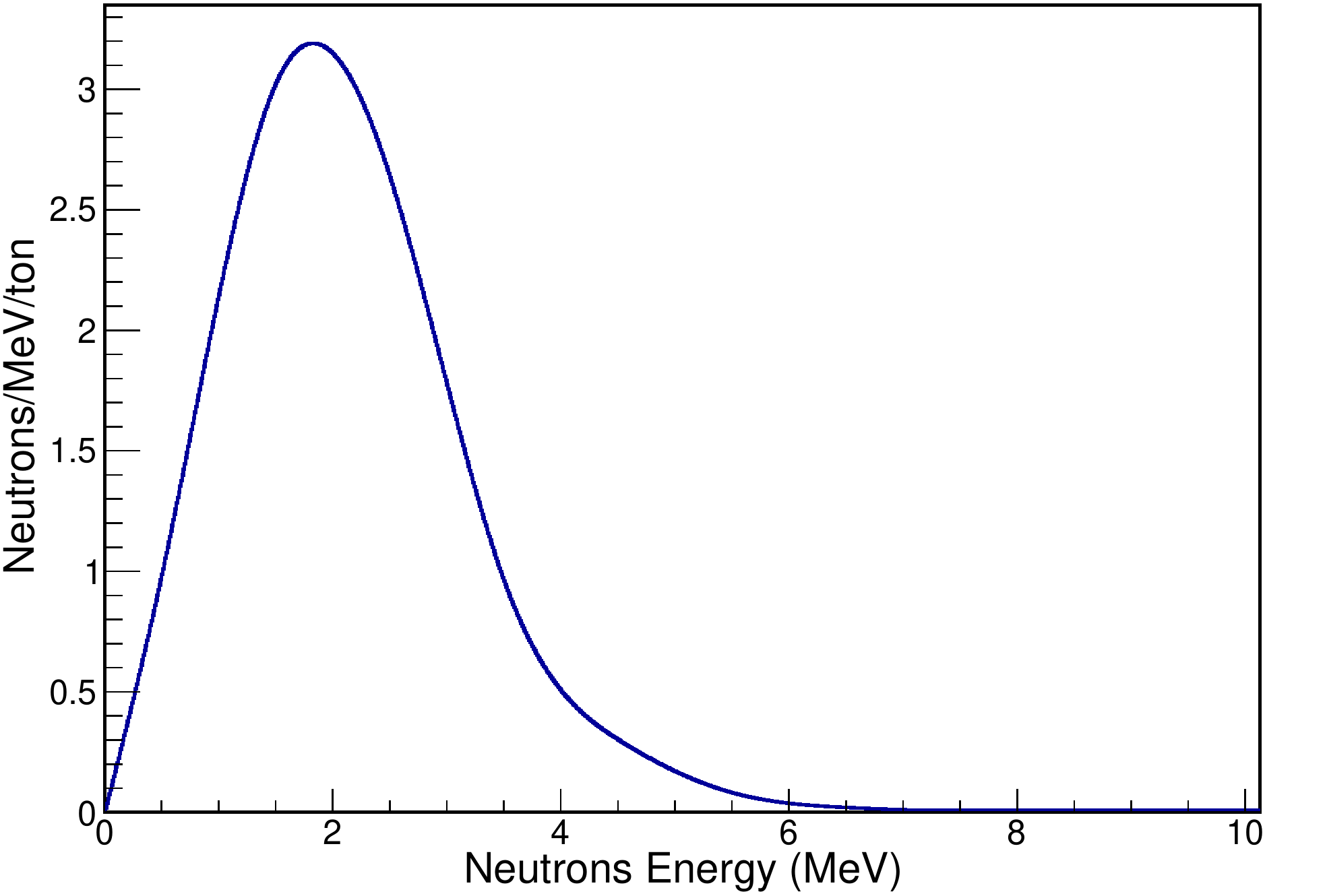}\hspace{1pc}%
\begin{minipage}[b]{14pc}\caption{\label{nuIn_spec_recSpec}Spectral distribution of neutrons from $\nu_e$ CC interactions with Xenon nuclei (see text)\cite{Xenon_myPaper}.}
\end{minipage}
\end{figure}

Supernova (SN) neutrinos undergoing coherent elastic neutrino-nucleus 
scattering (CE$\nu$NS)~\cite{cenns_or1_Freedman,cenns_or2_Freedman,cenns-detection} $\hspace{0.8mm}$may give rise to detectable number of nuclear recoil (NR) events  in future multi-ton scale liquid xenon scintillation detectors designed primarily for direct dark matter 
search~\cite{multi-ton-dm-detectors_Xe,multi-ton-dm-detectors_LZ,multi-ton-dm-detectors_Da}. In addition to the elastic scattering, the neutrinos from the SN may also undergo inelastic interaction with the detector nuclei resulting in production of neutrons within the detector. For instance, with $^{132}{\rm Xe}$ as the target nucleus, the electron type neutrinos ($\nu_e\hspace{0.5mm}$s) from the SN may undergo charged current (CC) interaction to produce an electron and a $^{132}{\rm Cs}$ nucleus in an excited state: 
\begin{equation}
\nu_e + {}^{132}_{54}\hspace{1mm}{\rm Xe}\rightarrow e^- + {}^{132}_{55}\hspace{1mm}{\rm Cs^*}\,\,.
\end{equation} 
The excited $^{132}{\rm Cs}$ nucleus may then decay through emission of one or more neutrons:
\begin{equation}
{}^{132}\hspace{1mm}{\rm Cs^*}\rightarrow{}^{132-X}\hspace{1mm}{\rm Cs}+ Xn \,\,,
\end{equation} 
where $X$ is the number of neutrons produced in the decay of the excited Cesium nucleus. The spectral distribution of neutrons produced in liquid xenon due to the above process for an 18$M_\odot$ progenitor SN burst at a distance of 1 kpc from the Earth, calculated using the neutrino flux given by the Basel-Darmstadt (BD) simulations of Ref. \cite{Fischer:2009af_BD_SN}, is shown in Fig. \ref{nuIn_spec_recSpec}. 
Elastic as well as inelastic scatterings of these neutrino induced neutrons ($\nu$I$n$) with the target nuclei would then give rise to further nuclear recoils in addition to those directly produced by the SN neutrinos through the CE$\nu$NS$\hspace{0.8mm}$ process. The nuclear recoil spectrum due to the CE$\nu$NS process can be calculated using the available analytical expression for the recoil spectrum. However, a proper calculation of the nuclear recoil spectrum due to (in general multiple) elastic and inelastic scattering of the neutrino-induced neutrons with the target nuclei requires a simulation of the neutron scattering process.

\section{Simulation of Nuclear Recoils}
In this section we discuss the simulation of the scattering of the supernova neutrino induced neutrons in a generic liquid xenon detector and the resulting xenon nuclear recoil spectrum, using the GEANT4\cite{Geant4Pap} simulation toolkit. For simplicity, here we consider only a single isotope of xenon, namely, $^{132}{\rm Xe}$, as the target nucleus for illustrating the results. Considering a density of $\sim2.953$ $\rm g/cm^3$, a cylindrical tank of both diameter and height of $\sim75.4$ cm has been chosen so as to accommodate 1 tonne of liquid xenon. The energy spectrum of neutrons produced by SN neutrinos through inelastic neutrino-nucleus scattering with $^{132}{\rm Xe}$ is found to extend up to $\sim$ 10 MeV with a peak at an energy of $\sim$ 2 MeV (see Fig. \ref{nuIn_spec_recSpec}). 
\begin{figure}[h]
\centering
\includegraphics[width=18pc]{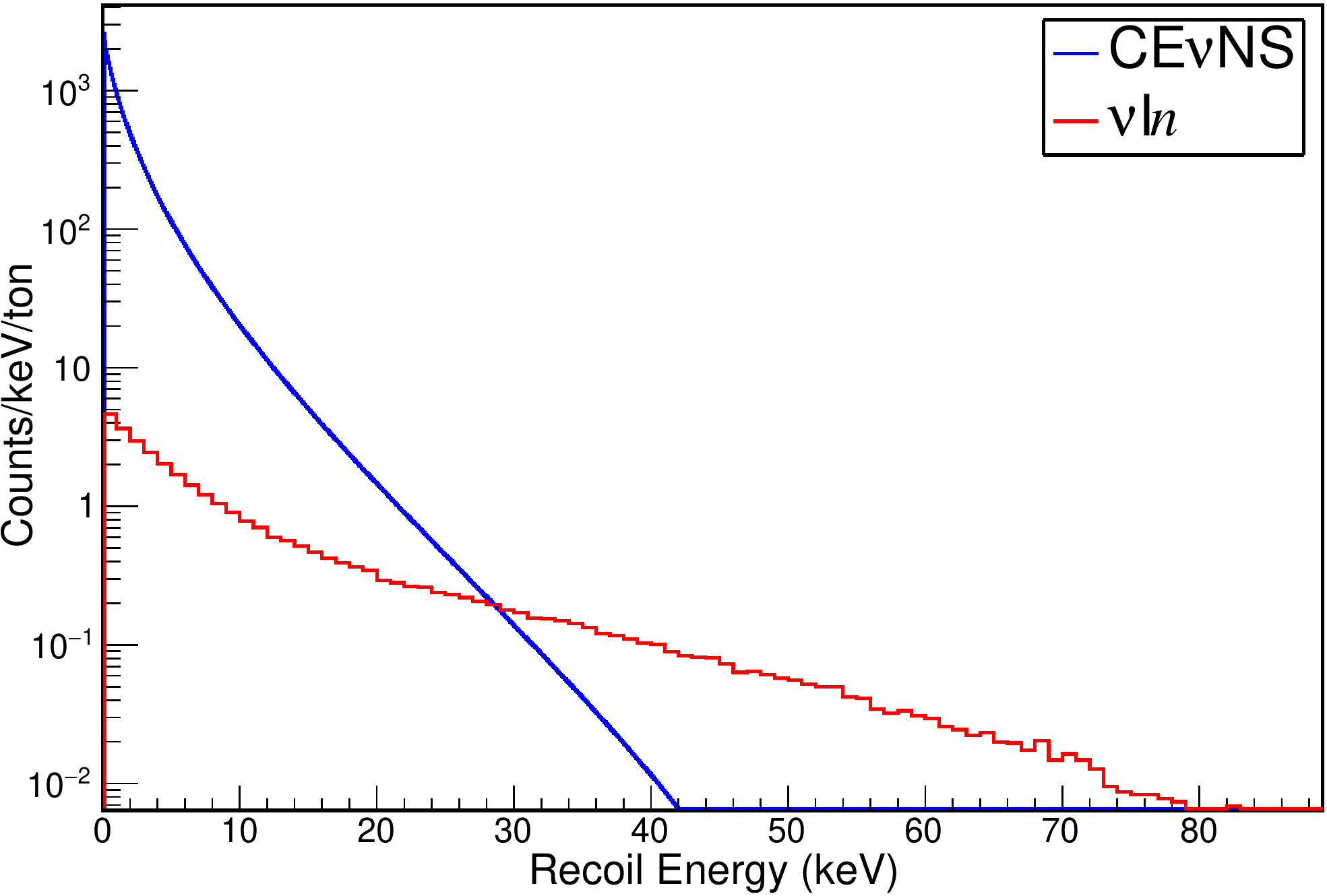}\hspace{1pc}%
\begin{minipage}[b]{14pc}\caption{\label{RecSpec_Xe}Differential nuclear recoil spectrum due to CE$\nu$NS and $\nu$I$n$ events\cite{Xenon_myPaper}.}
\end{minipage}
\end{figure}
At these energies, elastic scattering of the neutrons dominates the interaction of the neutrons with the target xenon nuclei. At each such scattering, the target xenon nucleus gets a recoil after the scattering event, with the neutron scattered off with a diminished energy. This process can take place multiple times during the passage of the neutron through the detector volume until the neutron exits the detector volume. Although the cross section is lower (compared to elastic scattering), the neutrons can also undergo inelastic scattering processes whereby, for example, a neutron may be absorbed by a $^{132}{\rm Xe}$ nucleus and is re-emitted along with a gamma ray, resulting in a small recoil of the nucleus. The emitted neutron can in turn undergo elastic or inelastic interaction with a $^{132}{\rm Xe}$ nucleus, and so on, until the neutron exits the detector volume. Although the probabilities of multiple scattering and inelastic scattering processes are lower than single elastic scattering events, the former processes together can give significant contribution to the overall nuclear recoil spectrum because of the higher number of nuclear recoils produced. We find that although the nuclear recoils due to CE$\nu$NS are more abundant at lower recoil energies, the nuclear recoils produced by the neutrino induced neutrons contribute dominantly to the overall nuclear recoil spectrum at recoil energies $\gtrsim$ 25 keV as shown in Fig. \ref{RecSpec_Xe}.

\section{Computation of Observable Signal}
Using the differential nuclear recoil scattering spectra produced through CE$\nu$NS and $\nu$I$n$, one can calculate the observable S1 and S2 signals in both these cases in a typical liquid Xenon based Time Projection Chamber (TPC) detector. The differential spectrum can be written as\cite{Lang_S1_S2} 
\begin{equation}
\frac{d^2R}{dS1~dS2} = \int dt_{\rm pb} dE_{\rm R}{\rm pdf}({\rm S1,S2}|E_{\rm R})\frac{d^2R}{dt_{\rm pb}~dE_{\rm R}}\,\,,
\end{equation}
where $t_{\rm pb}$ is the post-bounce time of the supernova burst event and the joint probability for getting a given S1 and a given S2 signal for a given recoil energy $E_{\rm R}$ is represented by pdf(S1,S2$|E_{\rm R}$). Following the detailed description in~\cite{Lang_S1_S2}, one can numerically compute the 
\begin{figure}[h]
	\centering
	\includegraphics[scale=0.38]{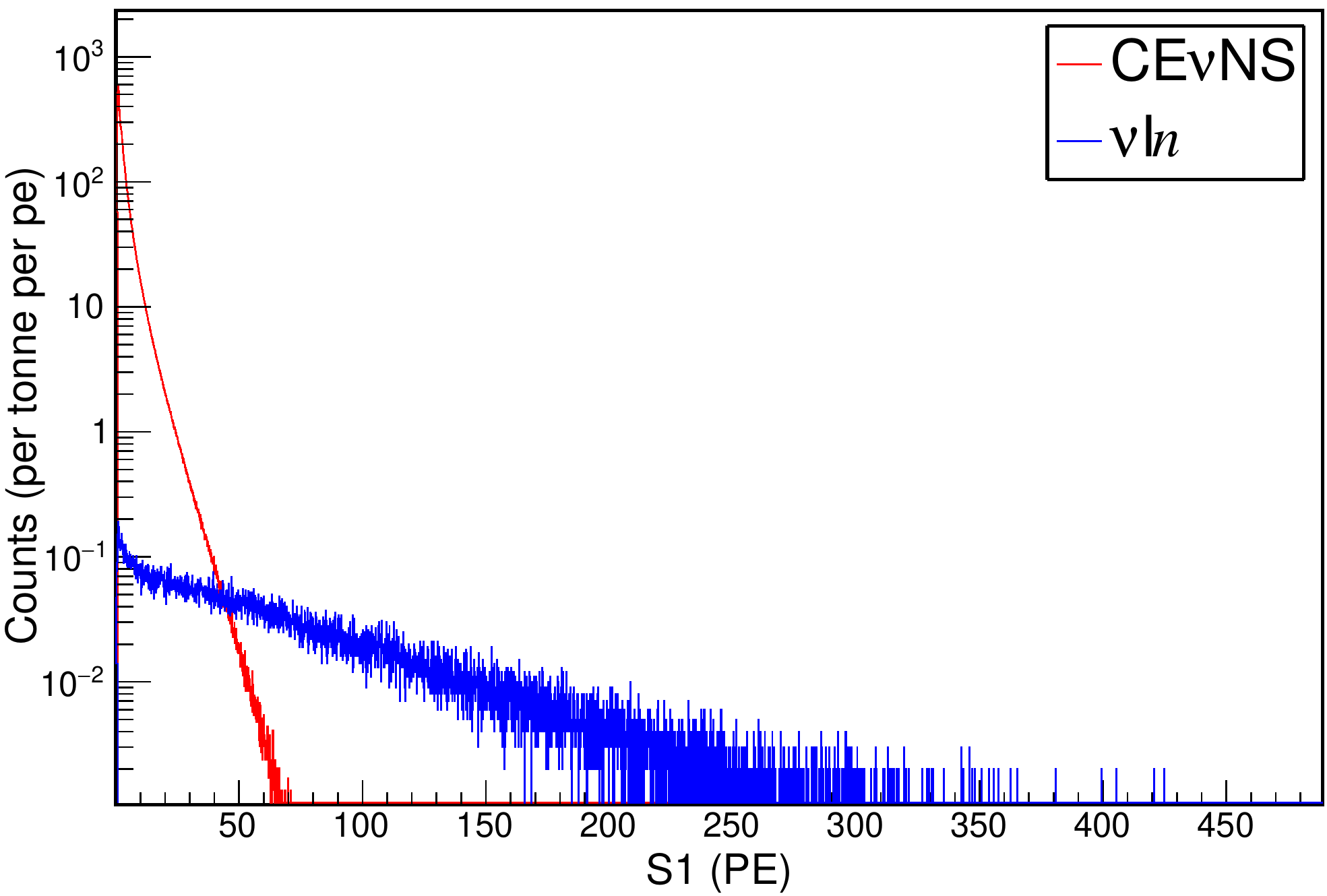}
	\hspace{0.5 mm}
	\includegraphics[scale=0.38]{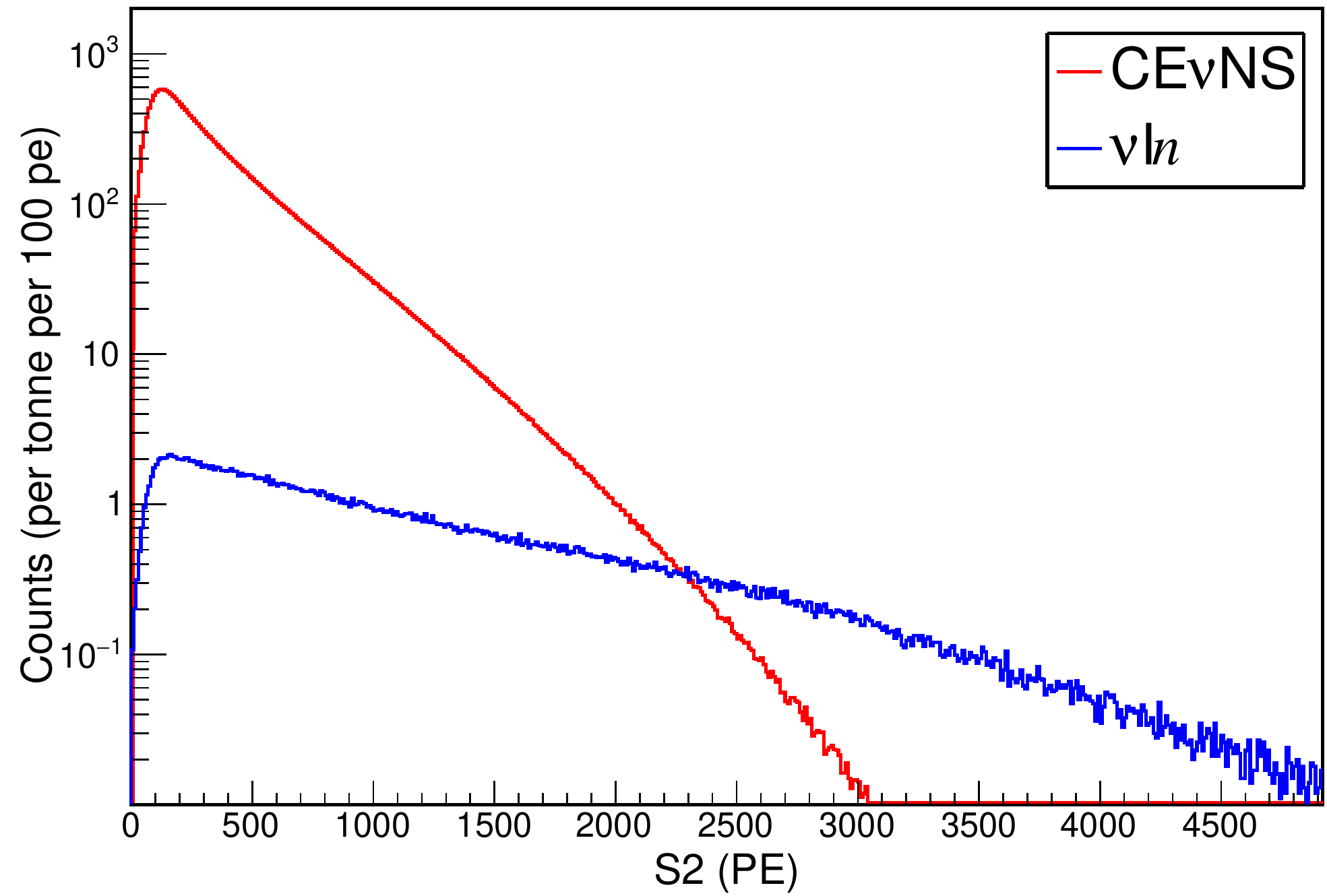}
	\caption{Differential spectrum for the S1 (left) and S2 (right) signals for CE$\nu$NS and $\nu$I$n$.}
	\label{S1S2_diff}
\end{figure}
S1 and S2 signals for both CE$\nu$NS and $\nu$I$n$. While each CE$\nu$NS event potentially produces one nuclear recoil (thereby producing a single S1 and a S2 signal), each neutrino induced neutron undergoing multiple scattering produces multiple S1 and S2 signals. In current experiments, operating with large liquid Xenon detectors, the typical sampling rate for the S1 signal is around 1 per 10 ns. This time scale is much bigger than the time taken by light to travel across the detector volume, given the current physical scale of detectors. Given the mean de-excitation lifetime of the excited xenon atoms, the width of a S1 pulse is around 27 ns\cite{Aprile_S1Width_Xenon} and therefore S1 signals originating from consecutive scattering events from the same neutron within say 50 ns of each other would overlap and be indistinguishable. Therefore, since the mean free time between two successive interactions of neutrons of typical energy of $\sim$ 2 MeV in liquid xenon is $\sim$ 6.5 ns, we see that S1 signals originating from events within 7 - 8 mean free paths would overlap and are merged into a single S1 signal. The S2 signals, on the other hand, are generated from the collision of the ionization electrons with the xenon atoms in the gas phase, the electrons originally being produced inside the liquid phase and drifted in the vertical ($z$) direction by a drift field with a typical drift velocity of $\sim2$ mm/$\mu$s. The width of the S2 signal is typically 1 - 2 $\mu$s\cite{XENON_S2}. Therefore, the S2 signals produced due to the arrival of ionization electrons at the gas phase, within $\sim 2$ $\mu$s of each other, would be merged into a single S2 signal.

\begin{figure}
    \centering
    \includegraphics[scale=0.38]{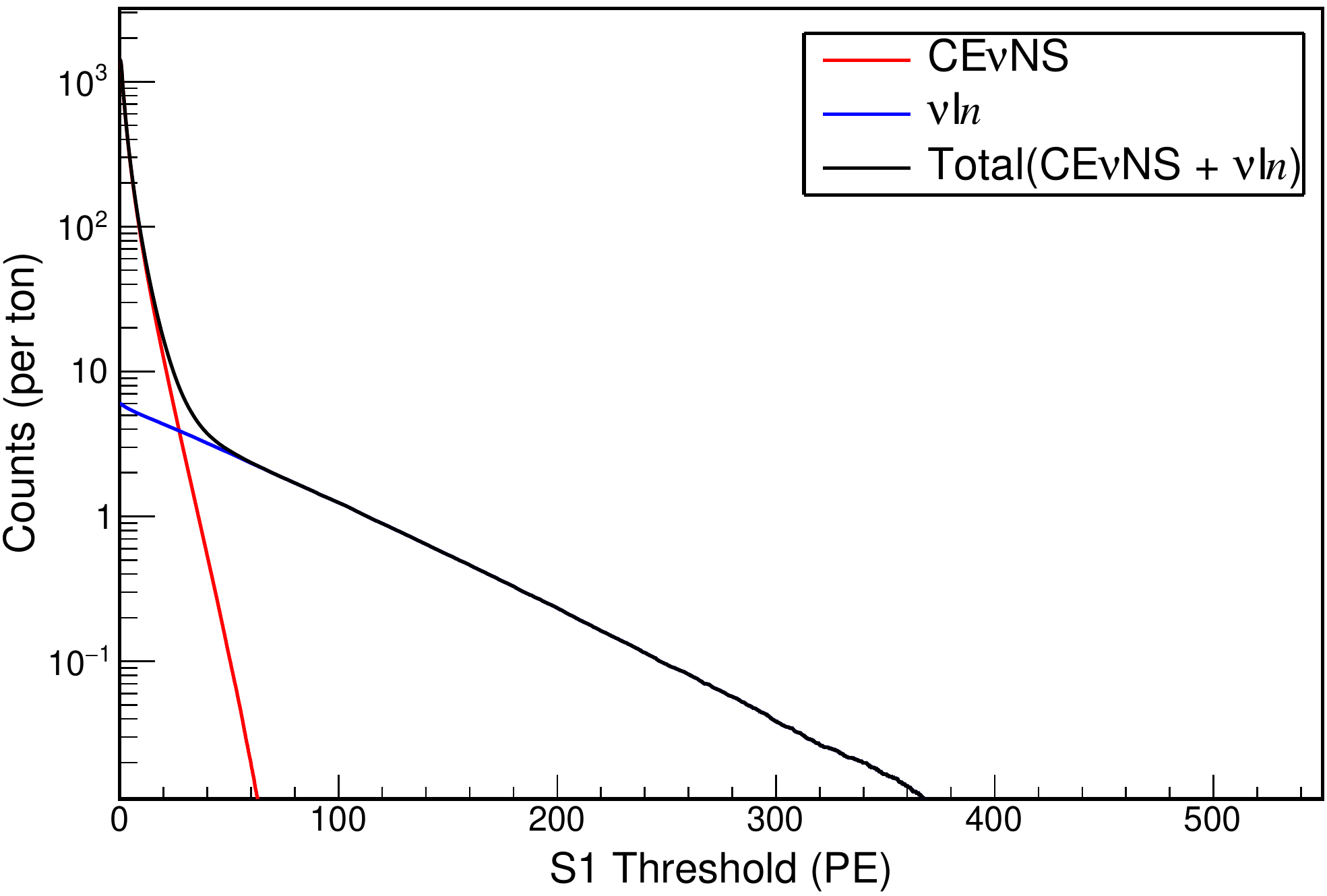}
	\hspace{0.5 mm}
	\includegraphics[scale=0.38]{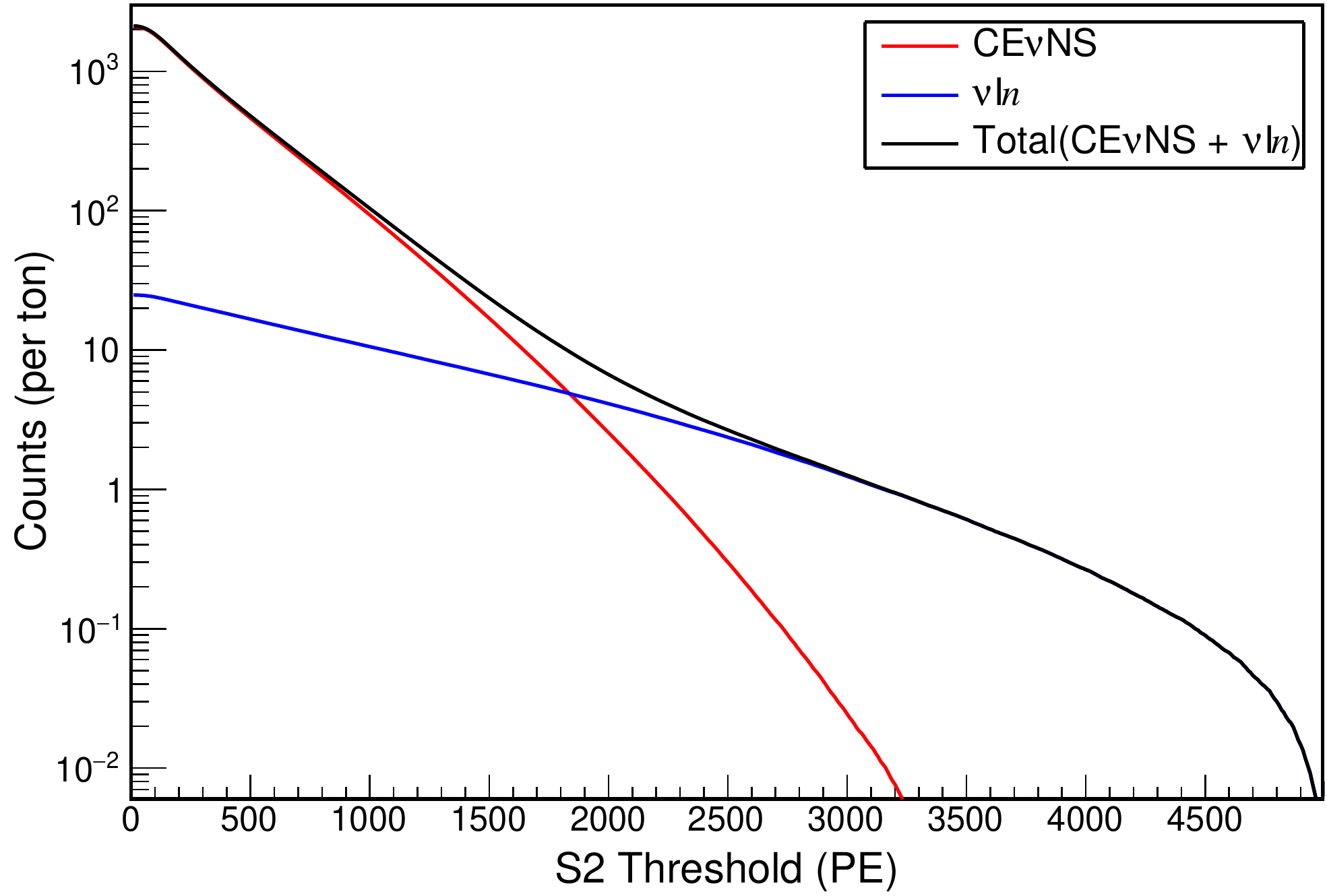}
	\caption{The integral S1 (left) and S2 (right) spectra as a function of detector threshold for the CE$\nu$NS and $\nu$I$n$.}
	\label{S1S2_integ}
\end{figure}
The differential and the integral spectra (as a function of detector threshold) of the S1 and S2 signals are shown in Fig. \ref{S1S2_diff} and Fig. \ref{S1S2_integ}, respectively. It is found that at low S1 and S2 thresholds, the CE$\nu$NS events dominate. However, at relatively large thresholds, 40 photo-electrons for S1 and 2500 photo-electrons for S2, the $\nu$I$n$ contributions to these two observable signals dominate over those due to  CE$\nu$NS. 

\section{Conclusion}
In this work we report the results of a simulation of the interaction of the neutrons, produced due to the CC interactions of SN electron neutrinos inside a typical liquid Xenon detector, with the xenon nuclei. We find that the $\nu$I$n$ contribution significantly modifies the recoil spectrum at large recoil energies and dominates over the CE$\nu$NS contribution at recoil energies $\gtrsim$25 keV. A MC simulation was used to compute the S1 and S2 signals from the nuclear recoil spectrum, and it was observed that at high detector thresholds, the contribution to the total integrated number of events from the $\nu$I$n$ scatterings dominates over those from CE$\nu$NS. It is interesting to note that while the nuclear recoil spectrum due to the CE$\nu$NS process receives contributions from all the species of neutrinos emitted in the SN burst, the $\nu$I$n$ contribution comes primarily from the $\nu_e$s from the SN burst event. Therefore, careful measurements of the full nuclear recoil spectrum, in both the S1 and S2 channels, due to SN burst events in future multi-ton scale liquid xenon dark matter detectors may provide useful information about the total SN explosion energy going into various neutrino flavors.

\section*{References}
\bibliography{Refs}
\end{document}